\documentclass[prb, twocolumn, showpacs, showkeys, superscriptaddress]{revtex4}%
\usepackage{amsfonts}
\usepackage{amsmath}
\usepackage{amssymb}
\usepackage{psfrag}
\usepackage{eucal}
\usepackage{units}
\usepackage{graphicx}

\usepackage[breaklinks=true]{hyperref}

\newcommand{\ev}{\frac{1}{4}}
\newcommand{\eh}{\frac{1}{2}}

\newcommand{\s}{\sigma}
\newcommand{\e}{\varepsilon}
\newcommand{\up}{\uparrow}
\newcommand{\down}{\downarrow}
\newcommand{\pd}{ {\phantom{\dagger}} }
\newcommand{\ii}{\mathrm{i}}
\newcommand{\Gc}{\mathcal{G}}

\begin{document}
\preprint{ }
\title[ ]{Nonequilibrium Cotunneling through a Three-Level Quantum Dot}
\author{S. Schmaus}
\affiliation{Institut f\"{u}r Theorie
der Kondensierten Materie, Universit\"{a}t Karlsruhe (TH), D-76128
Karlsruhe, Germany}
\author{V. Koerting}
\email[author to whom correspondence should be addressed:
]{verena.koerting@unibas.ch} 
\affiliation{Department of Physics, University of Basel,
Klingelbergstrasse 82, CH-4056 Basel, Switzerland}
\author{J. Paaske}
\affiliation{The Niels Bohr Institute \& Nano-Science Center,
University of Copenhagen, DK-2100, Copenhagen, Denmark}
\author{T. S. Jespersen}
\affiliation{The Niels Bohr Institute \& Nano-Science Center,
University of Copenhagen, DK-2100, Copenhagen, Denmark}
\author{J. Nyg{\aa}rd}
\affiliation{The Niels Bohr Institute \& Nano-Science Center,
University of Copenhagen, DK-2100, Copenhagen, Denmark}
\author{P. W\"{o}lfle}
\affiliation{Institut f\"{u}r Theorie der Kondensierten Materie,
Universit\"{a}t Karlsruhe, D-76128 Karlsruhe (TH), Germany}


\begin{abstract}
We calculate the nonlinear cotunneling conductance through a quantum
dot with 3 electrons occupying the three highest lying energy
levels. Starting from a 3-orbital Anderson model, we apply a
generalized Schrieffer-Wolff transformation to derive an effective
Kondo model for the system. Within this model we calculate the
nonequilibrium occupation numbers and the corresponding cotunneling
current to leading order in the exchange couplings. We identify the
inelastic cotunneling thresholds and their splittings with applied
magnetic field, and make a qualitative comparison to recent
experimental data on carbon nanotube and InAs quantum-wire
quantum dots.
Further predictions of the model such as
cascade resonances and a magnetic-field dependence of the orbital level
splitting are not yet observed but within reach of recent 
experimental work on carbon nanotube and InAs nanowire quantum dots.
\end{abstract}
\date{\today}
\maketitle


The Kondo effect has been observed in a number of different
quantum dot (QD), and single molecule
devices~\cite{Goldhaber:98,Cronenwett:98,Wiel:00,Nygaard:00,Sand:06}.
The effect is manifested as a sharp conductance peak at zero
bias-voltage, developing when the temperature is lowered beyond the
characteristic Kondo temperature. It relies on a spin-degenerate
ground-state on the quantum dot, which gives rise to logarithmically singular
spin-flip scattering of the conduction electrons traversing the dot.
Meanwhile, for quantum dots with sufficiently small level spacings
or slightly broken degeneracies, the Kondo-peak at zero bias voltage
can be flanked by two or more satellite steps or even peaks in the
nonlinear conductance. Such inelastic cotunneling features are often
seen in both GaAs\cite{Zumbuhl:04}, Carbon
nanotube~\cite{Babic:04,Paaske:06,Holm:08} (CNT) and
InAs-wire~\cite{Sand:06,Csonka:08} quantum dots, but in most cases
they are masked by charge-excitations which can be nearby in energy
and for this reason they have not received much attention.

Single-molecule transistors, on the other hand, exhibit a much
larger charging energy, $E_C$ ($\sim$ 100 meV instead of 5 meV, say,
for a typical quantum dot). At the same time, these molecular
systems often display a number of degeneracies which are weakly
broken once the molecule is contacted by source, and drain
electrodes, thus inducing splittings of the order of a few meV. Most
recently, this was seen in Refs.~\onlinecite{Osorio:07,Roch:08},
where junctions holding an OPV5, or a $C_{60}$ molecule,
respectively, showed a very clear singlet-triplet splitting on the
scale of 1 meV together with a charging energy of the order of 100
meV. This is a very convenient separation of energy scales which
raises the experimental resolution of inelastic cotunneling
phenomena to new standards. 
Nevertheless, given the simpler level-structure of most
conventional QD-devices, it is desirable to revisit and
understand the details of interorbital transitions better in
these systems. This is what we set out to do in the
present paper.

\begin{figure}[b]
\begin{center}
\includegraphics[width=0.9\linewidth]{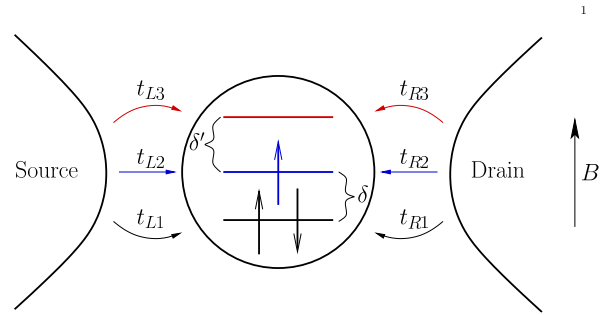}
\end{center}
\caption{Illustration of the setup. We consider the three
highest-lying levels of a quantum dot, separated by energies
$\delta$, and $\delta'$. Both orbitals are connected to source (left lead) and
drain (right lead) via 6 different tunnel couplings $t_{\alpha n}$. 
The dot is in the Coulomb blockade
regime with the three levels adjusted by a gate to accommodate
exactly 3 electrons.}
\label{model}
\end{figure}

\begin{figure}[t]
\begin{center}
\includegraphics*[width=0.8\linewidth,viewport=0 0 290 220,clip]{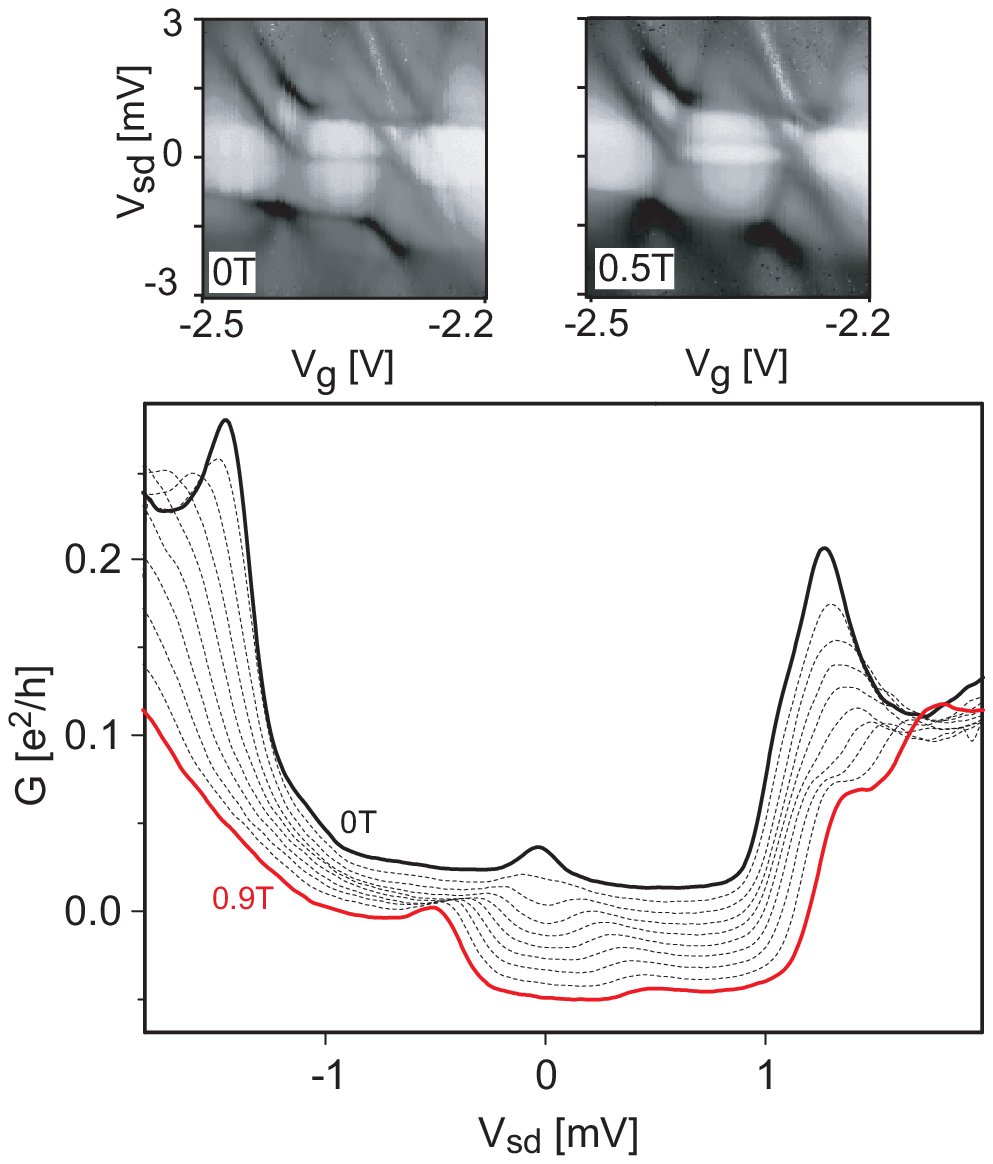}
\end{center}\vspace{-5mm}
\caption{(color online)
Differential conductance, $G$, as a function of bias
voltage, $V_{sd}$, for an InAs-wire based quantum dot at $T = 0.3$ K. The
data were taken at magnetic fields (perpendicular to the wire)
$B = 0$ (thick), 0.1 (dotted),
..., 0.9 T (red) and the curves were offset by 0.008 $e^2/h$ for
clarity. The data were taken for an odd occupied Coulomb diamond at
gate voltage $V_g = -2.35$ V \cite{Sand:06}.}
\label{fig:InAsdat}
\end{figure}

\begin{figure}[t]
\begin{center}
\includegraphics*[width=0.85\linewidth,viewport=0 10 290 220,clip]{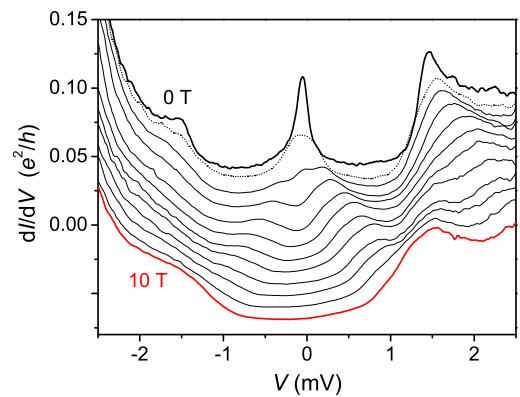}
\end{center}\vspace{-5mm}
\caption{(color online)
Differential conductance, $dI/dV$, as a function of bias
voltage, $V$, for a carbon nanotube quantum dot at $T = 0.08$ K. The
data were taken at magnetic fields (perpendicular to the tube)
$B = 0$ (thick), 0.1 (dotted), 1 (thin),
2, 3, ..., 9, 10 T (red) and the curves were offset by 0.008 $e^2/h$
for clarity. The data were taken for an odd occupied Coulomb diamond
at gate voltage $V_g = -4.96$ V \cite{Paaske:06}. (Note that at
finite magnetic fields features are broadened due to noise induced by
the magnet power supply).}
\label{fig:CNTdat}
\end{figure}

In the following, we study the case of a quantum dot occupied by an
odd number of electrons, featuring a spin-doublet groundstate,
giving rise to a zero-bias Kondo peak, but with additional orbitals/levels
leading to flanking inelastic cotunneling steps or peaks. The basic
3-orbital Anderson model is illustrated in Fig.~\ref{model} and will
be shown to host a variety of different I-V-characteristics
depending on the relative magnitudes of the 6 different
tunneling-amplitudes.
This 3-orbital model was briefly discussed by some of the present
authors in Ref.~\onlinecite{Sand:06} (cf.~inset in Fig.~4b), where
it was invoked to explain the relatively sharp peaks at finite bias
flanking a zero-bias Kondo-effect observed in an InAs quantum-wire
dot~\cite{Sand:06}. The measured nonlinear conductance curves for
varying applied magnetic fields are shown in Fig.~\ref{fig:InAsdat}.
This experiment constitutes one of the rare cases where such
side-peaks could actually be resolved.
As one further example, Fig.~\ref{fig:CNTdat} shows similar data recorded
on a single-walled carbon nanotube (CNT) quantum dot (QD). Both sets of
measurements show well-defined peaks which split into weak thresholds on
applying a magnetic field. Notice the different field-strengths needed in
the two experiments, reflecting the roughly 4 times larger spin-orbit
enhanced g-factor in InAs as compared to the CNT.

The bulk of this paper deals with leading order nonequilibrium cotunneling
for the 3-orbital Anderson model in the Kondo-regime. First we derive an
effective cotunneling, or Kondo model for a system with three electrons
distributed on the three orbitals. From this effective low-energy model we
then proceed to calculate the I-V characteristics to leading (second)
order in the cotunneling amplitude. In section III we discuss some of the
salient transport features of this system and in section IV we discuss the
CNT and the InAs data shown in Figs.~\ref{fig:InAsdat}
and \ref{fig:CNTdat}.
In this context we point out additional properties of such samples
which can be read off the dI-dV characteristic.

\section{Effective low-energy Kondo model}

The 3-orbital Anderson model corresponding to the setup in
Fig.~\ref{model} we write as
\begin{equation}
H=H_{\mathrm{lead}} + H_{\mathrm{dot}} + H_{\mathrm{tun}}\, .
\label{eq:AM}
\end{equation}
The leads are described by the non-interacting Hamiltonian
\begin{equation}
H_{\mathrm{lead}}=\sum_{\substack{\alpha=L,R\\\mathbf{k}\sigma}}(\varepsilon
_{\mathbf{k}}-\mu_{\alpha})c_{\alpha\mathbf{k}\sigma}^{\dagger}c_{\alpha
\mathbf{k}\sigma}^\pd
\end{equation}
\noindent where $\alpha=L,R$ labels the leads, assumed to be in
equilibrium at chemical potentials $\mu_{\alpha}$. The operator
$c_{\alpha\mathbf{k}\sigma }^{\dagger}$ creates an electron in lead
$\alpha$ of momentum $\mathbf{k}$ and spin $\sigma$. A simple
constant interaction model is used to describe the quantum dot
itself in terms of three non-degenerate orbitals with one common
Coulomb repulsion $U (\sim E_{C})$:
\begin{align}
H_{\mathrm{dot}}= &  \sum_{n\sigma}\varepsilon_{n\sigma}f_{n\sigma}^{\dagger
}f_{n\sigma}^\pd\nonumber\\
+ &  \frac{1}{2}U 
\sum_{n\sigma}\sum_{m\sigma^{\prime}}
f_{n\sigma}^{\dagger} f^\pd_{n\sigma}
\Big( f_{m\sigma^{\prime}}^{\dagger}f^\pd_{m\sigma^{\prime}} - 1 \Big) ,
\end{align}
where $n,m=1,2,3$ label the orbitals on the dot and
$f_{n\sigma}^{\dagger}$ creates an electron in orbital $n$ with spin
$\sigma$ and with energy
$\epsilon_{n\sigma}=\epsilon_{n}-\frac{1}{2}\sigma g\mu_{B}B$ where
$\epsilon_{2}\equiv\epsilon _{1}+\delta$, and
$\epsilon_{3}=\epsilon_{2}+\delta'$, with level-splittings denoted
by $\delta$ and $\delta'$. Finally, the tunneling Hamiltonian,
\begin{equation}
H_{\mathrm{tun}}=\sum_{\substack{\alpha=L,R\\n\sigma}}\left(  t_{\alpha
n}c_{\alpha\sigma}^{\dagger}f^\pd_{n\sigma}+h.c.\right)  ,\nonumber
\end{equation}
describes the coupling of the leads to the dot via six independent
tunneling amplitudes $t_{\alpha n}$, see illustration in
Fig.~\ref{model}. For later convenience, we have introduced the
local conduction electron operators
$c_{\alpha\sigma}^{\dagger}=
\sum_{\mathbf{k}}c_{\alpha\mathbf{k}\sigma}^{\dagger}$.

\begin{figure}[t]
  \centering
  \includegraphics[height=4cm,width = 0.3 \textwidth]{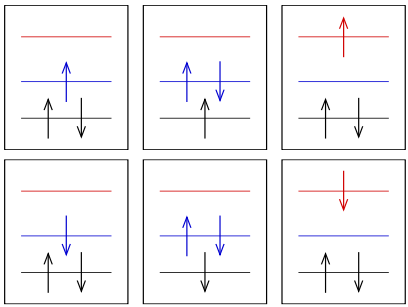}
  \caption{Illustration of the three lowest lying energy states that are taken
          into account in the effective cotunneling model of the 3-orbital
          Anderson model. From left to right: \textit{spin, hole}
          and \textit{particle} states.}
  \label{fig:plot2}
\end{figure}

We now restrict our attention to the Kondo-regime in which all three
orbitals are sufficiently narrow compared to the charging energy to
effectively suppress all charge-fluctuations, leaving the dot with a well
defined occupation of 3 electrons in three orbitals. Since the
experimental data in Figs.~\ref{fig:InAsdat} and \ref{fig:CNTdat} do not
resolve higher lying peaks corresponding to excited states with two
electrons in orbital 3, we shall simplify our model further by also
omitting these higher lying 3-particle states. In the experiments, the
cotunneling thresholds for these higher lying states must be comparable to
the charging-energy and are therefore masked by charge-fluctuations. As
mentioned earlier, this is a typical problem of the relatively large
quantum dots (compared to single-molecule junctions) which provide a
rather poor separation of energy-scales. Altogether, we are now left with
an effective low-energy Hilbert-space spanned by the six lowest lying
three-electron states:
\begin{align}
| s \s \rangle
&=f_{2\sigma}^{\dagger}f_{1\downarrow}^{\dagger}f_{1\uparrow}^{\dagger}
|\mathrm{vac}\rangle,\nonumber\\
| h \s \rangle
&=f_{2\downarrow}^{\dagger}f_{2\uparrow}^{\dagger}f_{1\sigma}^{\dagger}
|\mathrm{vac}\rangle,\\
| p \s \rangle
&=f_{3\sigma}^{\dagger}f_{1\downarrow}^{\dagger}f_{1\uparrow}^{\dagger}
|\mathrm{vac}\rangle.\nonumber
\end{align}
We label these three-body states by indices $a,b=\{s,h,p\}$ (i.e.
\{{\it spin, hole, particle}\}), together with the spin-index
$\sigma=\{\uparrow,\downarrow\}$. The notation is obvious from
the illustration of the states in Fig.~\ref{fig:plot2}.
The energies of these states are
\begin{align}
E_{s\sigma}&=2\varepsilon_{1}+\varepsilon_{2}+3U-\frac{1}{2}\sigma
g \mu_B B,\nonumber\\
E_{h\sigma}&=\varepsilon_{1}+ 2\varepsilon_{2}+3U-\frac{1}{2}\sigma
g \mu_B B,\\
E_{p\sigma}&= 2\varepsilon_{1}+\varepsilon_{3}+3U-\frac{1}{2}\sigma
g \mu_B B.\nonumber
\end{align}
In the case of equidistant energy levels $E_{h\sigma}=E_{p\sigma}$
and the two excited states, corresponding to respectively an
electron moved to orbital 3 or a hole moved to orbital 1, are
degenerate. The energy $\varepsilon_{2}$ is set by tuning the
voltage on a gate and here we choose it such that the
3-particle state will be lower in energy than both the 2, and the
4-particle states, that is:
\begin{align}
E_{2e^{-}}=2\varepsilon_{1}+U &  >E_{3e^-},\\
E_{4e^{-}}=2\varepsilon_{1}+2\varepsilon_{2}+6U &  >E_{3e^-}.\nonumber
\end{align}
In the following, we shall choose $\e_2 = - 5/2\ U$, which places
the system at the particle-hole symmetric point where $E_{4e^-} -
E_{3e^-} = E_{2e^-} - E_{3e^-}=U/2$, with $E_{3e^-} = E_{s\s}$ as
the ground state.

In order to eliminate charge-fluctuations from the 3-orbital
Anderson model (\ref{eq:AM}), we employ a generalized
Schrieffer-Wolff transformation~\cite{Schrieffer:66} which serves to
eliminate the tunneling-term, $H_{\mathrm{tun}}$, and retains only
terms of second order in the tunneling-amplitudes in the form of an
effective cotunneling, or Kondo Hamiltonian:
\begin{equation}
H_{\mathrm{eff}}=H_{\mathrm{lead}}+\sum_{a\sigma}E_{a\sigma}|a\sigma \rangle
\langle a\sigma|+ H_{\mathrm{int}}.
\end{equation}
The effective interaction now takes the form of a spin/orbital
exchange-term:
\begin{equation}
H_{\mathrm{int}}=\sum_{\substack{\alpha\beta,ab\\\s'\s}}\!\!J_{\alpha\beta}^{ab}
\left[\eh \mathbf{S}_{ab}\cdot\boldsymbol{\tau}_{\s'\s}
-\hat P_{ab}\delta_{\s'\s}\right]c_{\alpha\s'}^{\dagger}
c^\pd_{\beta\s},\label{eq:Hint}
\end{equation}
in terms of the vector of Pauli matrices, $\boldsymbol{\tau}$, the
spin-operator for the quantum-dot
\begin{equation}
\mathbf{S}_{ab}=\frac{1}{2} \sum_{\sigma\sigma^{\prime}} | a \sigma
\rangle \boldsymbol{\tau}_{\sigma\sigma^{\prime}} \langle b\sigma'
|,
\end{equation}
and the potential scattering term $\hat P_{ab} = \hat P_{ab}^\mathrm{inel} 
+ \hat P_{ab}^\mathrm{el}$
which consists of an inelastic scattering which involves a change in the
orbital state
\begin{align}
\hat P^\mathrm{inel}_{ab}=&\frac{1}{4} \big(\delta_{a, s}\delta_{b, p} 
   + \delta_{a, p} \delta_{b,s} 
   - \delta_{a, s} \delta_{b, h} - \delta_{a, h} \delta_{b, s} \big)
\nonumber \\
&\hspace{8mm}\times\sum_{\s'} | a \s' \rangle \langle b\, \s' |.
\end{align}
and the elastic scattering which occurs via empty levels,
i.e.~level $3$ for $| s \rangle$ and $| h \rangle$
and level $2$ for $| p \rangle$,
\begin{align}
\hat P^\mathrm{el}_{ab}=&\frac{1}{4} \delta_{a,b} \sum_{\tilde a}
\big(\delta_{\tilde a, s}\delta_{a, p}
+ \delta_{\tilde a, h}\delta_{a, p}  + \delta_{\tilde a, p} \delta_{a, s}
\big)
\nonumber \\
&\hspace{8mm}\times\sum_{\s'} | \tilde a \s' \rangle \langle \tilde a \s' |.
\end{align}
To lowest order a transition between the excited states 
$| p \rangle$ and $| h \rangle$ is not possible.

A constant energy offset arising in the Schrieffer-Wolff
transformation is neglected and so are 
further potential scattering terms. These lead only to a constant offset
in the differential conductance of the order of $\delta/U^2$, which can be
neglected in our qualitative study. Details of the calculation, including
the expression for the coupling functions $J_{\alpha\beta}^{ab}$, are
given in appendix~\ref{app:J}.

Note that when calculating the effective cotunneling amplitudes in the
appendix, we retain the differences in energy-denominators of the
different amplitudes, i.e. at this stage we do not make the approximation
that $U\gg\delta^{(\prime)}$. The according differences in amplitudes
expresses the broken orbital symmetry of the model, even with all six
tunneling-amplitudes being equal. Nevertheless, in order for the
Schrieffer-Wolff transformation to be meaningful, we must demand these
differences to be small, and in all our numerical calculations we
therefore choose a large charging energy ($U\sim 100\delta$) which
effectively makes all energy denominators equal. In other words, retaining
$\delta^{(\prime)}$ in the denominators of the cotunneling amplitudes
would require a more careful treatment of charge-fluctuations.

\section{Nonequilibrium perturbation theory}

When the bias voltage is large enough to populate the excited states
on the dot, these will no longer be thermally
occupied~\cite{Paaske:04a, Parcollet:02}. This effect is
incorporated in the Keldysh component Dyson equation expressed in
terms of nonequilibrium Green functions~\cite{Rammer:86, Haug:96}
with self-energies calculated to leading (second) order in the
effective cotunneling amplitudes or exchange couplings
$\nu_{F}J_{\alpha\beta}^{ab}\ll 1$. Following the approach taken in
Ref.~\onlinecite{Paaske:04a}, we employ a pseudo-fermion
representation for the dot-states, with operators defined by
$d^\dag_{a\s}|0\rangle\equiv|a\s\rangle$ and $\langle
0|d^\pd_{a\s}\equiv\langle a\s|$, and subject to the constraint
$\mathcal{Q} = \sum_{a,\s}d^\dag_{a\s}d^\pd_{a\s}=1$. The constraint is enforced
with the aid of a Lagrange multiplier $\lambda$, included as an
additional term, $H_{\lambda}=\lambda
\sum_{a,\s}d^\dag_{a\s}d^\pd_{a\s}$, in the
Hamiltonian~\cite{Abrikosov:65}. 
The exact projection to the physical Hilbert space is
effected by taking the limit $\lambda \to \infty$.
We shall need the non-equilibrium Green's functions:
\begin{align}
\mathcal{G}_{ab,\sigma\sigma^{\prime}}(\tau,\tau^{\prime}) &  =-\mathrm{i}
\left\langle T_{\mathrm{C_{K}}} \big( d^\pd_{a\sigma}(\tau)d_{b\sigma^{\prime}}^{\dagger
}(\tau^{\prime}) \big) \right\rangle \\
G_{\alpha\beta,\sigma\sigma^{\prime}}(\tau,\tau^{\prime}) &  =-\mathrm{i}
\left\langle T_{\mathrm{C_{K}}} \big( c^\pd_{\alpha\sigma}(\tau)c_{\beta\sigma^{\prime}
}^{\dagger}(\tau^{\prime}) \big) \right\rangle
\end{align}
with $T_{C_{K}}$ being the time ordering operator along the Keldysh
contour. Calligraphic letters denote pseudo-fermion, italic letters
conduction electron Green's functions.

The retarded and advanced Green's functions can be calculated like in the
equilibrium case directly from the Dyson equation. The pseudo-fermion
spectral function is obtained from the imaginary part of the retarded
Green's function,
\begin{equation}
\mathcal{A}_{aa,\sigma}=-2\ {\rm Im}\left[\frac{1}{\omega-\omega
_{a\sigma}-\Sigma_{aa,\sigma}^{R}}\right]  \label{A},
\end{equation}
and takes the approximate form of a Lorentzian at the resonance
frequency $\omega_{a\sigma}=\varepsilon_{a\sigma}+\lambda$ and of
width $\Gamma_{aa,\sigma }=-2\ {\rm Im}[\Sigma_{aa,\sigma}^{R}]$.
Since the spectral function appears in later evaluations only in
convolution with functions that vary on a larger energy scale than
$\Gamma_{aa, \sigma}$, we approximate it 
by a simple delta-function:
\begin{equation}
\mathcal{A}_{aa,\sigma}(\omega)=2\pi\delta(\omega-\omega_{a\sigma}).
\end{equation}
The lesser function is found from the quantum Boltzmann equation (QBE), or
generalized Kadanoff-Baym equation:
\begin{equation}
\Gamma_{aa,\sigma}(\omega)\mathcal{G}^{<}_{aa,\sigma}(\omega)=
\mathcal{A}_{aa,\sigma}(\omega)\Sigma^{<}_{aa,\sigma}(\omega).
\label{QBE}
\end{equation}
\noindent Within the delta-function (quasiparticle) approximation
for the spectral
function, this equation can be solved using the following ansatz:
\begin{equation}
\mathcal{G}_{aa,\sigma}^{<}(\omega)=\mathrm{i}n_{a\sigma}\mathcal{A}_{a\sigma
}(\omega),\label{Gkl}
\end{equation}
through which the QBE takes the form of a simple rate-equation.

The conduction electrons are assumed to remain in thermal equilibrium and are
therefore characterized simply by their respective chemical potentials
together with a simple flat-band approximation for the momentum-summed
spectral functions, or local conduction electron density of states (DOS)
at the contact,
\begin{equation}
A(\omega)=2\pi \nu_{F}\theta(D-|\omega|),
\end{equation}
in terms of the DOS at the Fermi surface, $\nu_{F}$, and half
bandwidth $D$.

\subsection{Nonequilibrium occupation numbers}

In writing the QBE in Eq.~\eqref{QBE}, we have tacitly assumed the
pseudo-fermion self-energies to be diagonal in both spin, and
orbital indices. Neglecting spin-orbit interactions 
spin is a conserved quantum number and the
self-energy will therefore be diagonal in this index. This does not
hold for the orbital quantum number $a$ and off-diagonal terms,
$\Sigma_{ab,\s}$, can arise. Nevertheless, in the regime studied
here for which the level splittings $\delta,\delta'$ are assumed to
be much larger than level broadenings, $\Gamma_{a}$, energy
conservation suppresses such off-diagonal terms and the self energy
can safely be assumed to be diagonal. For an example in which
off-diagonal contributions remain important we refer the interested
reader to Ref.~\onlinecite{Koerting:08}.

Thus neglecting off-diagonal self-energies and using Eq.~\eqref{Gkl}, the
QBE in Eq.~\eqref{QBE} takes the following form:
\begin{equation}
n_{a\sigma}=\mathrm{i}\frac{\Sigma_{a\sigma}^{<}(\omega_{a\sigma})}
{\Gamma_{a\sigma}(\omega_{a\sigma})}\label{sce}.
\end{equation}
The self energy $\Sigma_{a\s}^<$ itself depends on the occupation numbers
and this therefore constitutes a set of six coupled equations for six
unknown occupation numbers. These equations are underdetermined and should
therefore be solved together with the constraint
$\mathcal{Q}=\sum_{a\sigma}n_{a\sigma}=1$. More details of the actual
calculation of pseudo-fermion self-energies are given in
appendix~\ref{app:selfen}.

\subsection{Cotunneling current}

We obtain the current operator directly from the time-derivative of
the density operator at the contact in the left lead, say 
(see Ref.~\onlinecite{Paaske:04a} and references therein). With
$n_{L}=\sum_{\sigma}c_{L\sigma}^{\dagger}c^\pd_{L\sigma}$, one finds
\begin{align}
j_{L} &= e\frac{d n_{L}}{dt}
=\frac{\mathrm{i}e}{\hbar}\left[  n_{L},H_\mathrm{int}\right]  \nonumber\\
&  =\sum_{\alpha,ab,\sigma'\sigma}
\bigg\{\frac{1}{2}\mathbf{S}_{ab}\cdot\boldsymbol{\tau}_{\sigma\sigma^{\prime}}
\left(J_{LR}^{ab}c_{L\sigma'}^{\dagger}c^\pd_{R\sigma}
    - J_{RL}^{ab} c_{R\sigma'}^{\dagger}c^\pd_{L\sigma}\right)  \nonumber\\
&\qquad+{\hat P}_{ab} \delta_{\s', \s}
\left( J_{LR}^{ab}c_{L\sigma}^{\dagger}c^\pd_{R\sigma}
     - J_{RL}^{ab}c_{R\sigma}^{\dagger}c^\pd_{L\sigma}\right)
\bigg\}.
\end{align}
In terms of the correlation functions,
\begin{align}
D_{\alpha\beta}^{\mathrm{spin}}(\tau,\tau^{\prime}) & =-\mathrm{i}\langle
T_{{C}_K}\sum_{ab, \s'\s}J_{\alpha\beta}^{ab}\mathbf{S}_{ab}(\tau)
\eh \boldsymbol{\tau}_{\s'\s} 
c^\dag_{\alpha \s'}(\tau') c^\pd_{\beta \s}(\tau')
\rangle,\nonumber\\
D_{\alpha\beta, \s}^{\mathrm{pot}}(\tau,\tau^{\prime}) & =-\mathrm{i}\langle
T_{{C}_K}\sum_{ab} J_{\alpha\beta}^{ab} {\hat P}_{ab}(\tau)
c_{\alpha\sigma}^{\dagger}(\tau^{\prime})c^\pd_{\beta\sigma}
(\tau^{\prime})\rangle,\nonumber\\
D_{LR}^{\mathrm{tot}}(\tau,\tau') &  =
D_{LR}^{\mathrm{spin}}(\tau,\tau') + D_{LR}^{\mathrm{pot}}(\tau,\tau'),
\end{align}
the expectation value of the current is given in lowest order
in the couplings $J_{LR}^{ab}$ by
\begin{align}
& \left\langle j_{L}\right\rangle = - \frac{4\pi e}{h}{\rm Re}\left[
D_{LR}^{\mathrm{tot},>}(\tau,\tau)\right]
\nonumber\\ =&
   \Big( \frac{\pi}{2} \Big)^2 \frac{e}{h} \nu_F^2
   \sum_{ab,\sigma\sigma^{\prime}}
   \Big\{ 
     J_{LR}^{ab} J_{RL}^{ba}
\nonumber \\ & \times 
   \Big( 2 \boldsymbol{\tau}_{\s'\s}\boldsymbol{\tau}_{\s\s'}
        + \sum_{c = \{ p, h\}} \big(
            \delta_{a, s} \delta_{b, c} + \delta_{a, c} \delta_{b, s} \big)
   \Big)
\nonumber \\ & \times
  \Big( n_{a\s} Y(\e_{b\s'} -\e_{a\s} - e V) 
      - n_{b\s'} Y(\e_{a\s} -\e_{b\s'} + e V) 
  \Big)
\nonumber \\ & 
+ 
  eV\ J_{LR}^{aa} J_{RL}^{aa} 
  \sum_{\tilde a} n_{\tilde a, \s}\ 
  \big( \delta_{a, p} \delta_{\tilde a, s}
      + \delta_{a, p} \delta_{\tilde a, h}
      + \delta_{a, s} \delta_{\tilde a, p} \big)
  \Big\}
  \label{eq:current}
\end{align}
where $Y(x) = x\ n_B(x)$ and 
$n_B(x) = 1/(\exp[x/2T] - 1)$ denotes the Bose distribution.
The sum goes over all possible states $a, b = \{ s, h, p \}$, which are
weighted according to the nonequilibrium occupation numbers
$n_{b\sigma'}$.

\section{Discussion of results}

Tuning the six different tunneling amplitudes, the two different
level-spacings and external magnetic field allows for a large variety of
cotunneling I-V characteristics. With an eye towards the experiments
mentioned in the introduction we shall discuss a few of the salient
features. Since we restrict our calculations to leading order perturbation
theory, we do not address the interesting question of Kondo-correlations
in neither the zero-bias peak, nor any of the finite-bias conductance peaks.

\subsection{Occupation numbers}

\begin{figure}[t]
\begin{center}
\includegraphics[width=0.9\columnwidth]{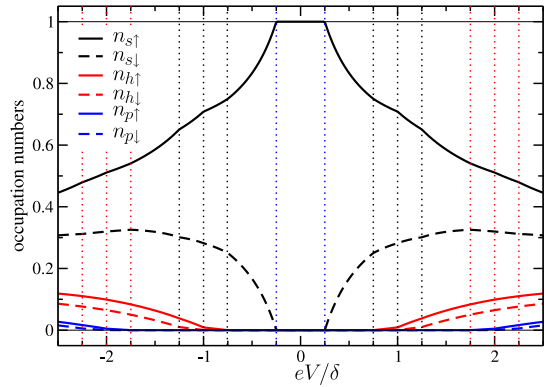}
\end{center}
\caption{(color online) 
Occupation numbers of the \textit{spin, hole} and
\textit{particle} states in the $3$-level Anderson model for
symmetric coupling to the leads, $t_{\alpha n} = 0.1 \sqrt{U/\nu_F}$. 
For a clear separation of energy scales the temperature is chosen small
$T/\delta = 0.001$ and the levels are clearly separated
$\delta'/\delta = 2$. The dashed lines indicate the transition
energies $\{\pm g \mu_B B,\delta, \delta \pm g \mu_B B, \delta', \delta'
\pm g \mu_B B \}$ in $\{$blue, black, red$\}$
for a finite magnetic field of $B/\delta = 0.25$.
}
\label{fig:ocnum}
\end{figure}

Results for the occupation numbers are shown in
Fig.~\ref{fig:ocnum}. For convenience we perform the calculations
with a small thermal smearing. Nevertheless, the temperature is
chosen far smaller than any other energy scale in the problem
($T=0.001\delta$ throughout) and all plots can be thought of as
corresponding to $T=0$. For finite magnetic field and zero
bias-voltage the ground state, $|{s\up}\rangle$, is occupied with
probability one. As illustrated in Fig.~\ref{fig:ocnum}, the other
states become populated for larger voltages and in the limit $V \to
\infty$ all states will be equally occupied.
Inter-orbital transitions play a role as soon as the voltage becomes
larger than $\delta$ or $\delta'$, which is the energy needed to
excite respectively an electron into a higher-lying level, or a hole into
a lower-lying level. For equidistant levels ($\delta=\delta'$) these
excitations are identical and we find $n_{h \sigma} = n_{p \sigma}$.

\begin{figure}[t]
  \centering
  \includegraphics[width = 0.9 \columnwidth]{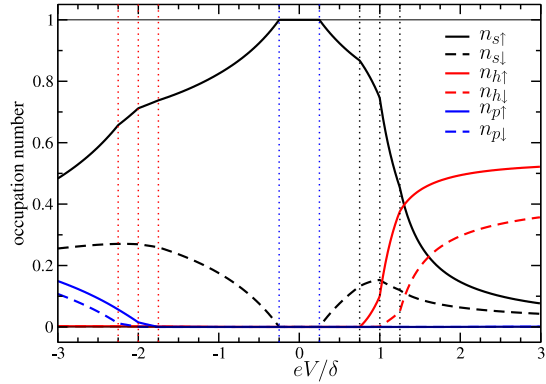}
  \caption{(color online)
Occupation numbers of the \textit{spin, hole}
and \textit{particle} states in the $3$-level Anderson model for
non-symmetric coupling to the leads, i.e.~$t_{L2}/t_{Rn} = 3$,
$t_{L3}/t_{Rn} = 1$ and $t_{L1}/t_{Rn} = 1/3$ where 
$t_{Rn} = 0.1 \sqrt{U/\nu_F}$.
With the same finite
magnetic field $B/\delta = 0.25$, level splitting
$\delta'/\delta = 2$ and temperature $T/\delta = 0.01$
as in Fig.~\ref{fig:ocnum}
we observe an occupation inversion around $eV
\approx \delta + B$ since $n_{h\up}$ becomes higher
occupied than the ground state $n_{s\up}$. This blocking of
transitions and the \textit{cascade} effect
appearing at $\delta^{(')}- g\mu_BB$ are discussed in the main text.
Gridlines refer to various thresholds: $\pm g \mu_B B$ (blue),
$\{\delta, \delta \pm g \mu_B B\}$ (black),
and $\{- \delta', - \delta' \pm g \mu_B B\}$ (red).}
  \label{fig:ocnumasym}
\end{figure}

In Fig.~\ref{fig:ocnumasym} we show the same occupation numbers as
in Fig.~\ref{fig:ocnum}, but now calculated for
different tunnel-couplings to the left lead. Again, the ground-state is
partially depleted with increasing voltage, but due to the
asymmetric couplings a clear asymmetry in bias is observed. For
negative bias, only the $p$-states are being populated, due to a
strong $|s\rangle\to |p\rangle$ (cotunneling) amplitude proportional to
$t_{L2}t_{R3}=3$. The $|s\rangle\to |h\rangle$ amplitude is proportional to
$t_{L1}t_{R2}=1/3$ and therefore only hardly pumped at all. For
positive bias, the situation is reversed since 
$\langle p|H_\mathrm{int}|s\rangle\propto t_{L3}t_{R2}=1$ is now much smaller than
$\langle h|H_\mathrm{int}|s\rangle\propto t_{R1}t_{L2}=3$, whereby it is
$n_{h}$ which rises with voltage. Notice also that since $\langle
h|H_\mathrm{int}|p\rangle=0$ there are no direct transitions between the
two excited states, whereby an incipient occupation of $|h\rangle$,
say, will be accompanied by a deletion of $|s\rangle$ and thereby
further prohibit the pumping of $|p\rangle$.

For the asymmetric tunnel-couplings chosen for
Fig.~\ref{fig:ocnumasym}, one can also observe a {\it population
inversion} at positive bias. That is, for $eV\gtrsim\delta+g\mu_BB$,
$n_{h\up}$ increases rapidly and stays larger than the
ground-state occupation $n_{s\up}$ for an extended bias-range. The
positive bias demands electrons to predominantly jump from the left
lead onto the dot and from the dot into the right lead. Therefore,
$t_{L2}\gg t_{L1}$ will make the $|s\rangle\to|h\rangle$ much more
likely than the $|h\rangle\to|s\rangle$ transition at positive bias
and the dot gets stuck in the excited state $|h\rangle$. As we shall
argue later, this excited state does not sustain as high a current
as the ground state and therefore this population inversion will in
fact lead to regions of negative differential conductance (NDC).

The 'direct' threshold for populating $|h,\up\rangle$, say, is given
by $eV=\delta$. Nevertheless, from Fig.~\ref{fig:ocnumasym} it is
seen that $n_{h\up}$ starts growing already at $e V=\delta - g \mu_B
B$. This slightly lower 'indirect' threshold is due to a {\it
cascade} effect in which $|s,\down\rangle$ is populated for
$V>g\mu_{B}B$ and from there the threshold to $|h,\up\rangle$ is now
only $\delta-g\mu_{B}B$ instead of $\delta$. Further cascade effects
are observed at $eV =\delta$ (instead of the direct threshold
$\delta+g\mu_{B}B$) which is the threshold for the transition
$|s,\down\rangle\to|h,\down\rangle$. A similar effect for
$n_{p,\sigma}$ is observed at negative bias, with indirect
thresholds already at $eV=-(\delta'-g\mu_{B}B)$ and $eV=-\delta'$.
Notice that these effects are just barely visible for symmetric
couplings as in Fig.~\ref{fig:ocnum}. As we shall see in the next
subsection, these \textit{cascade} effects show up as small steps in the
nonlinear conductance.

\subsection{Cotunneling conductance}

\begin{figure}[t]
\begin{center}
\includegraphics[width=0.9 \columnwidth]{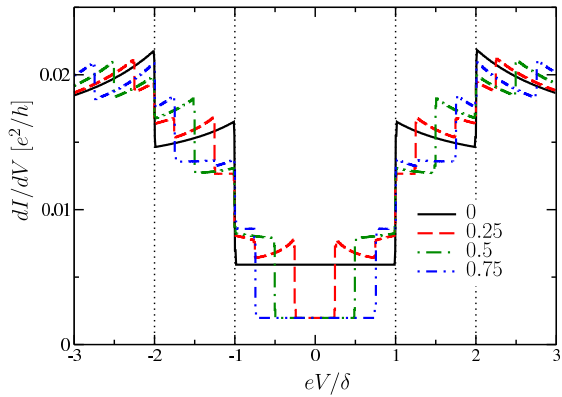}
\end{center}
\caption{(color online)
Dependence of the differential conductance $dI/dV$ on the
applied bias voltage in units of $eV/\delta$ for various strengths of
the magnetic field $g\mu_{B}B/\delta = 0, 0.25, 0.5, 0.75$. The
quantum dot is coupled symmetrically to the lead with the parameters
given in Fig.~\ref{fig:ocnum}. The kinks at $\pm (\delta - g\mu_{B}B)$ and
$\pm (\delta' - g\mu_{B}B)$ originate from the occupation of excited
states due to the \textit{cascade} effect.}
\label{fig:didvB}
\end{figure}
In Fig.~\ref{fig:didvB} we show the differential conductance for the
same parameter set as in Fig.~\ref{fig:ocnum}. As expected, we find
a step in $dI/dV$ whenever a new level enters the voltage window.
Furthermore, as pointed out in
Refs.~\onlinecite{Wegewijs:01,Paaske:04a}, the voltage-dependence of
the nonequilibrium occupation numbers promotes these conductance
steps to pointed cusps. Since we are limiting our calculations to
second order perturbation theory, these plots do not take into
account the possible Kondo-enhancement of these cusps. Nevertheless,
the tendency is known from previous calculations: each cusp in the
differential conductance to second order is logarithmically
enhanced~\cite{Paaske:04a} and these logarithmic divergences are
contained roughly by the inverse life-time of the excited state
involved in the relevant inelastic cotunneling
process~\cite{Paaske:04b, Rosch}.

\subsubsection{Cascade induced side-peaks}
Even for the symmetric couplings in Fig.~\ref{fig:didvB}, we observe
extra steps reflecting the aforementioned \textit{cascade} effect. For
example, the red (dashed) curve with  $g\mu_{B}B=0.25\delta$
shows extra steps both at $eV=\delta-g\mu_{B}B$
($|s,\down\rangle\to|h,\up\rangle$) and at $eV=\delta'-g\mu_{B}B$
($|s,\down\rangle\to|p,\up\rangle$). Notice that the \textit{cascade} effect
is only visible for $B \not= 0$, and disappears when $2\, g\mu_{B}B >
\delta (\delta')$ since $n_{h, \up} (n_{p, \up})$ is then occupied
before $n_{s,\down}$.
\begin{figure}[b]
  \centering
  \includegraphics*[width=0.5\columnwidth]{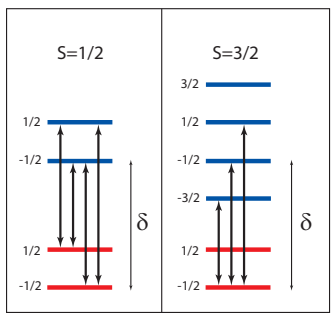}
  \caption{
  Comparison of direct and indirect transitions with a doublet
  groundstate and an excited doublet or quartet, respectively, at finite
  magnetic field. Cascade effect provides an indirect transition in the
  former, at energy $\delta-g\mu_{B}B$, which matches a direct transition
  in the latter.} \label{fig:spectr}
\end{figure}

More investigations are necessary in clarifying to what extent these
cascade-features in the conductance will be enhanced by the Kondo-effect, but
even a small additional step as seen in Fig.~\ref{fig:didvB} can have
important bearings for interpreting experiments in which the internal dot,
or molecule states are not known in advance. Judging from the magnetic
field dependence, the three transitions at
$eV={\delta-g\mu_{B}B,\delta,\delta+g\mu_{B}B}$ seen in
Fig.~\ref{fig:didvB} (imagine for a moment that $\delta'\gg\delta$) could
in fact be misinterpreted as direct transitions between a spin-doublet
ground state and an excited S=3/2 state, as illustrated by the
energy-diagrams in Fig.~\ref{fig:spectr}.
\footnote{The inherent spin-$3/2$ state of the model, 
if each level is occupied by each one electron, is a higher order excitation
with an energy of $\delta + \delta'$ above the ground state
and is therefore in experiments mostly masked by charge excitations.} 
This example shows that a
spin-spectrum read off from B-dependent inelastic cotunneling-lines in a
{\it diamond-plot} ($dI/dV$ vs. $V_{g}$ and $V_{sd}$) should be interpreted
with some care, especially when enhancing the lines by plotting higher
derivatives of the current.

\subsubsection{Negative differential conductance}

\begin{figure}[t]
  \centering
  \includegraphics[width=0.9 \columnwidth]{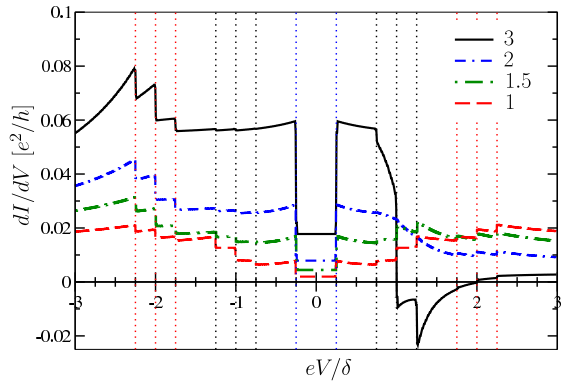}
  \caption{(color online)
Differential conductance $dI/dV$ of the
non-symmetric 3-level Anderson model with $3$ electron for the
parameters given in Fig.~\ref{fig:ocnumasym}. 
The four different set
of parameters correspond to the ratio of 
$t_{L2}/t_{Rn} = \{ 3, 2, 1.5, 1 \}$ with 
$t_{L1}/t_{Rn} = 1/(t_{L2}/t_{Rn}) = \{ 1/3, 1/2, 2/3, 1 \}$, respectively.
The parameter set with $t_{L2}/t_{Rn} = 3$ and $t_{L1}/t_{Rn} = 1/3$
corresponding to Fig.~\ref{fig:ocnumasym} 
shows NDC at the voltage values corresponding to
an inversion of occupation probabilities.}
  \label{fig:didvasym}
\end{figure}
In Fig.~\ref{fig:didvasym} we plot the conductance for the parameters used
in Fig.~\ref{fig:ocnumasym} and with varying values for the
tunneling amplitudes from the left lead to orbitals 1 and 2, respectively.
As in Fig.~\ref{fig:ocnumasym}, we observe a
pronounced asymmetry in bias-voltage, and with the largest difference in
couplings to orbitals 1 and 2 (black/solid curve), we observe a sudden drop to
negative differential conductance for $eV\gtrsim\delta$. As mentioned
before, the NDC is driven by the population inversion between $|s\rangle$
and $|h\rangle$. As argued in the previous section, the positive bias
drives the dot into the excited state $|h\rangle$, and since (for these
parameters) this state leads to a lower current than $|s\rangle$, the
current decreases and we observe the NDC. The poor cotunneling across the
dot in the state $|h\rangle$ is a simple consequence of the double
occupancy of the well-coupled orbital 2, which demands that orbital 2 is
emptied to the right lead before it can be filled from the left, thus
making use of the strong coupling $t_{L2}=3$. In contrast, starting in
$|s\rangle$ one can also fill in an electron in orbital 2 from the left
lead first and then tunnel out into the right lead from either orbital 2
or from orbital 1, thus allowing more 
current-carrying tunneling processes involving $t_{L2}=3$.

Thus a sufficiently strong difference in coupling to two of the orbitals
will cause a population inversion with a concomitant decrease in the
current as bias is increased. For intermediate coupling asymmetry, we see
from the other curves in Fig.~\ref{fig:didvasym} that the NDC disappears
while an asymmetry in bias remains.

\section{Discussion of experiments}

\subsection{Experiments on Carbon Nanotube}

The data shown in Fig.~\ref{fig:CNTdat} were recorded on the same sample as was
discussed in Ref.~\onlinecite{Paaske:06}. In fact, this is the neighboring
charge-state to the even number occupied state studied in that work. In
Ref.~\onlinecite{Paaske:06}, some of the present authors obtained a very
gratifying fit to the inelastic cotunneling line reflecting a strong
transition from singlet ground- to an excited triplet state. In particular,
it was argued that the very sharp cotunneling line could only arise from
the joint effect of nonequilibrium pumping (finite-bias occupations) of
the triplet state and substantial logarithmic enhancements from
the nonequilibrium Kondo-effect.

\begin{figure}[t]
  \centering
  \includegraphics[width = 0.9 \columnwidth]{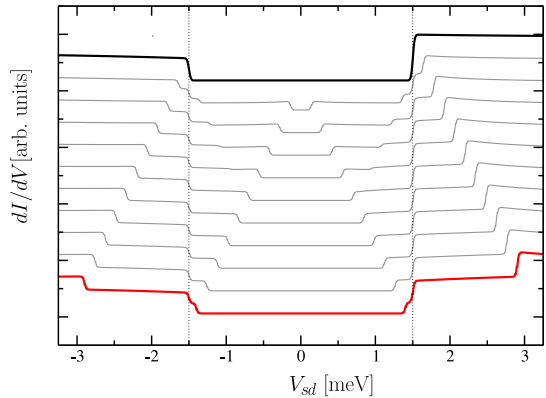}
  \caption{
  Modeling of the CNT QD analyzed in Ref.~\onlinecite{Paaske:06} as
  measured in the neighboring diamond, i.e.~Fig.~\ref{fig:CNTdat}, where
  orbital scattering is observed at $\delta = 1.5$ meV. Further parameters
  of the plot are $\{ t_{L1}, t_{L2}, t_{R1}, t_{R2} \} = \{ 0.032, 0.028,
  0.108, 0.063 \} \sqrt{U/\nu_F}$, $B = 0, 0.14, \ldots 1.4\ $meV, and $T =
  81$mK $\sim 0.008$meV. The conductance is given in arbitrary units
  and the curves are offset for clarity.}
  \label{fig:CNT}
\end{figure}

As already stated, we shall not embark on higher order perturbation theory
in this paper, which means that we should not expect to obtain any
quantitative agreement with the data in Fig.~\ref{fig:CNTdat}.
Nevertheless, since basically all parameters were fixed by the fit in
Ref.~\onlinecite{Paaske:06}, we shall assume that these remain largely
unchanged when passing on to the neighboring charge-state and use them as
input for a numerical evaluation of the nonlinear conductance there. The result
is shown in Fig.~\ref{fig:CNT} and the gross features such as bias-voltage
asymmetry and B-field splittings match those of Fig.~\ref{fig:CNTdat}
quite well, albeit with a complete lack of sharp finite-bias peaks, as
expected. The parameters from Ref.~\onlinecite{Paaske:06} were tunnel
couplings of the two lowest lying levels of $\{ t_{L1}, t_{L2}, t_{R1},
t_{R2} \} = \{ 0.032, 0.028, 0.108, 0.063 \} \sqrt{U/\nu_F}$ (see Fig.~4
in Ref.~\onlinecite{Paaske:06}), an orbital splitting of $\delta \approx
1.5$ meV, a charging energy of $U \approx 3.0$ meV, and a higher lying
level at $\delta^{\prime}=\Delta-\delta\approx 3.1$ meV, 
$\Delta$ being the single-particle level spacing in the nanotube. Since
$\delta^{\prime}\gtrsim U$, charge-excitations set in before orbital 3 is reached
and this latter orbital is therefore not included in the calculation.

\subsection{Experiments on InAs-nanowires}

The data shown in Fig.~\ref{fig:InAsdat} constitute the first observation
of Kondo-effect in InAs-nanowire quantum dots and were discussed by some
of the present authors in Ref.~\onlinecite{Sand:06}. Already in that paper the
3-orbital model was invoked to explain the finite-bias peaks flanking the
zero-bias Kondo peak and, as mentioned in the introduction, this
experiment was the main motivation for this more detailed investigation of
that model.

\begin{figure}[t]
  \centering\vspace*{-5mm}
  \includegraphics*[width=\columnwidth]{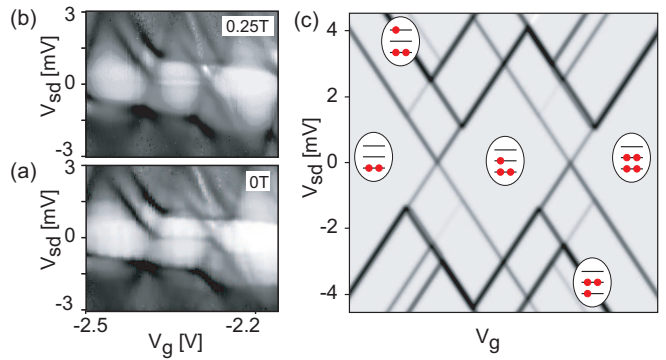}
\caption{(a) and (b) Plot of $dI/dV$ vs. gate, and
bias-voltage, for the InAs-wire quantum dot with $B = 0$ T and $0.25$ T,
respectively. Strong tunnel-coupling
perturb the Coulomb-blockade diamond and particularly strong
sequential tunneling lines are clearly present (thick dark lines) in
the region with a mixture of 2 and 3 electrons on the dot. 
c) Calculation of
the sequential tunneling lines from solving semiclassical
rate-equations (conductance in arbitrary units) bringing out some of 
the main lines and indicating the states which are
involved in the dominant transitions and which are involved
in the cotunneling for $N = 3$. Parameters are
$\delta=1.4$ meV, $\delta^{\prime}=1.1$ meV, $T=300$mK, $U=3$ meV,
$t_{L1}=t_{R1}=t_{L3}=t_{R3}=1$, $t_{L2}=0.4$, and $t_{R2}=2.5$.}
\label{fig:InAsExpSum}
\end{figure}

Apart from the zero-bias Kondo peak, it is the sharp finite-bias
peaks which dominate these data: A very strong peak close to
$eV=-1.4\ $ meV together with a somewhat weaker peak close to $eV=1.1\ $
meV (cf.~Fig.~\ref{fig:InAsdat}). This
asymmetry in bias-voltage was not addressed in
Ref.~\onlinecite{Sand:06}, but as we shall argue it can readily be
understood in terms of an asymmetry in tunneling amplitudes of
orbital 2 to source and drain. The excited state giving the strong
sequential tunneling line in the upper left corner of 
Fig.~\ref{fig:InAsExpSum} (a) is ascribed to the excited
3-particle state with one electron in orbital 3
instead of orbital 2. This line connects to the inelastic cotunneling line at
positive bias inside the N=3 diamond ($|s\rangle \to |p\rangle$ transition). 
From this we infer that $\delta^{\prime}\approx 1.1\ $meV. Notice that a
sequential tunneling line corresponding to a transition from the N=2
ground-state to the N=3 $|h\rangle$-state is only possible to higher order
in the tunneling amplitudes and is therefore strongly suppressed in
the experimental data. The inelastic cotunneling line at negative
bias can now be ascribed to an (N=3) $|s\rangle\to |h\rangle$ 
transition at energy $\delta\approx 1.4\ $meV. 
This is again consistent with the lower left
corner of Fig.~\ref{fig:InAsExpSum} (a) showing that this line connects
to a very strong sequential tunneling line which must correspond to
a transition from the N=4 ground-state to the N=3 $|h\rangle$-state. At this
point it is the (N=3) $|p\rangle$-state which is not directly coupled to the
N=4 ground-state and hence suppressed in the data.
In Fig.~\ref{fig:InAsExpSum} (c), we show the result of solving a set
of semiclassical rate-equations. The dark lines correspond to high
conductance (arbitrary units) due to sequential tunneling only. We do not
attempt a $6$ parameter fit of Fig.~\ref{fig:InAsExpSum} (a),
and tunneling amplitudes are therefore chosen with a very simple
asymmetry, assuming that only orbital $2$ is coupled in a special way. This 
choice of parameters captures some, but not all, of the gross features of the
data in Fig.~\ref{fig:InAsExpSum} (a).

\begin{figure}[t]
  \centering
  \includegraphics[width= 0.9\columnwidth]{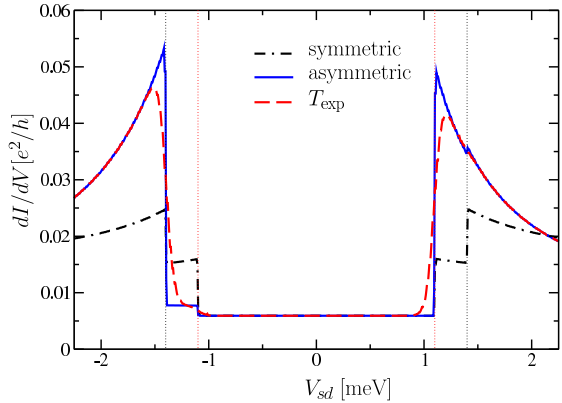}
  \caption{(color online)
  Plot of $dI/dV$ vs. bias voltage, $V$, calculated within second
  order perturbation theory. Parameters are chosen so as to approximately
  match the voltage asymmetry brought out by the experiment (compare
  Fig.~\ref{fig:InAsExpSum}). Parameters are $B=0$,
  $\delta=1.4$ meV, $\delta^{\prime}=1.1$ meV and $T = 0.001$ meV 
  (blue/solid curve). 
  The black (dashed-dotted)
  curve corresponds to all equal tunneling amplitudes and 
  the red (dashed) curve is
  the same as the blue (solid) but with temperature corresponding to the
  experiment ($T_\mathrm{exp}= 300$mK$\sim 0.025$meV).}
  \label{fig:InAsAsym}
\end{figure}

The question remains why there is only one and not two inelastic cotunneling 
peaks at both positive and negative bias. As demonstrated by the
calculation shown in Fig.~\ref{fig:InAsAsym}, however, the observed
asymmetry in bias-voltage can be understood quite simply as orbital
2 being asymmetrically coupled to the source and drain electrodes.
Notice that it is the nearly equidistant levels in the InAs-wire
quantum dot which prompts us to incorporate the effects of both $h$
and $p$ excited states. 
For asymmetric couplings the $\delta' = 1.1$meV and $\delta = 1.4$meV 
peaks in Fig.~\ref{fig:InAsAsym} dominate for positive and negative voltages, 
respectively, in contrast to symmetric
coupling where both can be observed. The small steps 
observed for zero temperatue (blue/solid curve) are washed out
for large temperature smearing (red/dashed curve) 
and the nonlinear conductance is now asymmetric with respect to voltage.
For a larger difference in the effective
$\delta$ and $\delta^{\prime}$, as is typical for carbon nanotube
quantum dots where  $\delta$ would be the subband-splitting
and $\delta^{\prime}$ would be of the order of the single-particle
level-spacing ($\Delta = \delta' + \delta$), the $|p\rangle$
state would most often be masked by charge-fluctuations, i.e.
$\delta^{\prime}$ will often be comparable to the charging-energy,
$E_C$.

\begin{figure}[t]
  \centering
  \includegraphics[width=0.9\columnwidth]{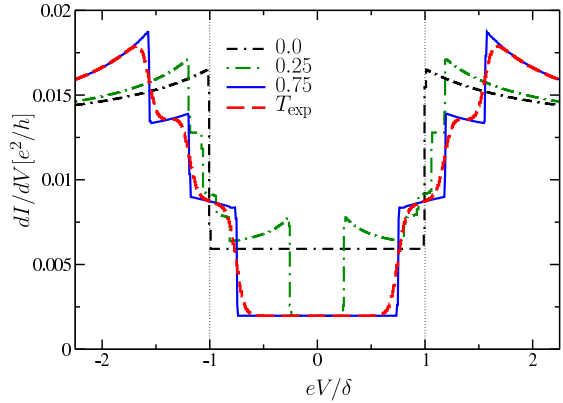}
  \caption{(color online)
  Plot of $dI/dV$ vs.~bias voltage, $V$, calculated within second
  order perturbation theory for symmetric coupling to the leads
  and neglecting the particle state ($\delta' > E_C$). 
  The $g$-factors are different for the two involved levels: $g_2 = 2 g_1$.
  The orbital peak moves towards higher
  voltages such as $\sim (1 - g_1/g_2) g_2\mu_BB/2$
  with increasing magnetic field $g_2 \mu_B B/\delta= \{0, 0.25, 0.75\}$. 
  The red (dashed) curve at $g_2 \mu_B B/\delta = 0.75$
  illustrates that the shift is also observable
  at a temperature corresponding to the
  experiment ($T_\mathrm{exp}= 300$mK$\sim 0.025$ meV).}
  \label{fig:InAsgfact}
\end{figure}

As demonstrated recently in the experiment by Csonka {\it et
al.}~\cite{Csonka:08}, the strong spin-orbit coupling in InAs can
give rise to very different g-factors for neighboring quantum dot
orbitals or energy-levels. In that experiment this was brought out
particularly clear by a modulation of the orbitals by an additional
top-gate. Fixing the top-gate voltage and adjusting the back-gate,
it was shown that two neighboring levels could have g-factors of 1.9
and 10, respectively. It is interesting to note that such a
difference in g-factors would give rise to an apparent
$B$-dependence of the level-spacings. In our model, we would expect
the following $B$-dependent inelastic cotunneling thresholds:
\begin{align*}
{\tilde \delta}(B)&=E_{h\uparrow}-E_{s\uparrow}
=\delta+(g_{2}-g_{1})\mu_{B}B/2,\\
{\tilde \delta}^{\prime}(B)&=E_{p\uparrow}-E_{s\uparrow}
=\delta^{\prime}+(g_{2}-g_{3})\mu_{B}B/2.
\end{align*}
In Ref.~\onlinecite{Sand:06}, a detailed investigation of the
splitting of both the zero-bias Kondo peak and the
finite(positive)-bias cotunneling peak with magnetic field indeed
revealed two different g-factors. Interpreting the finite-bias
cotunneling as above, we can thus infer from
Ref.~\onlinecite{Sand:06} that orbital 2 has $g_{2}=7.7$ and orbital
3 has $g_{3}=8.5$. This difference is rather small and would cause a
largely negligible shift of roughly 0.02 meV of $\delta^{\prime}$ at
largest applied fields (0.9 Tesla). In Fig.~\ref{fig:InAsgfact} we
show the result of a calculation with slightly different g-factors
by a factor of $2$.
This is the cause of the slight shift of the main kink in the blue (solid)
curve to a value above $eV=\delta$. The observation of such
$B$-dependent cotunneling thresholds thus provides an interesting
consistency check on the difference in g-factors of two neighboring
orbitals.

In closing, we note that the diamond-plot in the upper left panel of
Fig.~\ref{fig:InAsExpSum} has a few strong-coupling irregularities
which we do not address in this paper. First of all, the inelastic
cotunneling lines display a slight dependence on gate-voltage. This
is a feature which was investigated in detail for a carbon nanotube
quantum dot in Ref.~\onlinecite{Holm:08}, where it was explained in
terms of second order tunneling-renormalization with energy
denominators depending on gate-voltage. Second, the inelastic
cotunneling line at positive bias, which we assigned to the excited
$p$-state, appears to continue straight through the sequential
tunneling region mixing N=3 and N=4 states. At this point we merely
speculate that this is related to a very strongly coupled orbital 3,
which might allow the partially populated N=4 states to conduct via
virtual transitions to the N=5 state. How, and if, this would work
out in detail will be assessed in future work.

\section{Conclusions}

In this paper we have analyzed the inelastic cotunneling
spectroscopy of a system with three electrons distributed among
three orbitals. Starting from the 3-orbital Anderson model in the
Coulomb blockade regime, we have projected out charge-fluctuations
and derived an effective cotunneling (Kondo) model describing
elastic as well as inelastic tunneling induced transitions between
different orbital and spin states of the impurity or quantum dot in
the N=3 charge-state.
By treating all tunnel couplings independently we can study
asymmetric behavior inherent to most of the experimental samples.

All calculations were performed within second order nonequilibrium
perturbation theory, and the bias dependent nonequilibrium
occupations of the impurity-states were shown to give rise to
marked cusps at voltages matching the different transitions. We
demonstrate how a bias induced {\it cascade} in the occupations can
lead to indirect transitions between two excited states.
Furthermore, we find that certain asymmetries in the tunneling
amplitudes of the three orbitals can lead to a {\it population
inversion}, which in turn can be accompanied by lines of negative
differential ({\it cotunneling}) conductance (NDC).

We have revisited the carbon nanotube data from
Ref.~\onlinecite{Paaske:06} with focus on an odd occupied
charge-state. Importing the tunnel-couplings inferred from
Ref.~\onlinecite{Paaske:06}, we confirm the 
asymmetry in the conductance strength at positive and negative bias. 
These parameters lead to steps at finite-bias at the orbital
transitions, and we believe that the experimentally observed peaks 
are mainly due to logarithmic enhancements from a nonequilibrium Kondo-effect.

We have also analyzed the InAs-wire data from
Ref.~\onlinecite{Sand:06} in greater detail and provided an
explanation of the bias-asymmetry in peak-{\it positions}. This was
understood as a slight difference in the two level-spacings together
with an asymmetry in the tunnel-coupling of orbital 2 to source and
drain electrodes. This asymmetry in couplings was roughly consistent
with the overall stability diagram for the N=3 diamond, as confirmed
by a semiclassical rate-equation calculation. Finally, we have
pointed to an additional consequence of strong spin-orbit coupling
in the InAs-wires, namely the apparent $B$-dependence of the
level-spacing caused by different g-factors for different orbitals.
In the present experiment this difference was too small to give a
noticeable effect, but with differences as quoted in
Ref.~\onlinecite{Csonka:08} it should be a pronounced
effect~\footnote{From private communications with S. Csonka we have
learned that the relevant inelastic cotunneling lines were not
resolved by the bias-range explored in that experiment and therefore
the effect could not be checked in this experiment}. 
Possible further influence of strong spin-orbit coupling on the cotunneling
spectroscopy of inelastic spin and orbital transitions constitutes
an interesting problem on its own right and will be addressed in a
separate publication.

\section*{Acknowledgments}

We thank M.~Mason, L.~DiCarlo, C.~M.~Marcus (Harvard University), M.~Aagesen,
and C.~B.~S\o rensen (Niels Bohr Institute) for
contributions to the experiments.
Furthermore we like to thank S.~Csonka, M.~Trif, D.~Stepanenko, V.~Golovach
and B.~Trauzettel for valuable discussions.

This work has been supported
by the DFG-Center for Functional Nanostructures
(CFN) at the University of Karlsruhe 
(S.Sch.~and P.~W.), the Institute for Nanotechnology,
Research Center Karlsruhe (P.~W.), the Danish
Agency for Science, Technology and Innovation and
the Danish Natural Science Research Council  (J.~P., T.~S.~J., J.~N.),
the Swiss NSF and the NCCR Nanoscience (V.~K.).

\setcounter{secnumdepth}{1}
\begin{appendix}
\section{Schrieffer-Wolff transformation and Coupling Constants}
\label{app:J}

In this work we concentrate on the physics inside a Coulomb diamond
where the 3-level quantum dot is occupied by exactly three
electrons. All virtual processes to two or four electron states can
be treated perturbatively and lead to a Kondo-like interaction and
potential scattering terms.

Using a Schrieffer-Wolff transformation~\cite{Schrieffer:66} we find
to second order in the tunneling Hamiltonian $H_{tun}$ the
interaction Hamiltonian
\begin{align*}
  H_\mathrm{int} &=
\sum_{\substack{\alpha, \beta; n, m \\ \s, \s'}}
\Big\{  t_{\alpha m} t^*_{\beta n}
  c^\dag_{\alpha \s} f^\pd_{m\s} \frac{1}{E_{a\s} - E_{4e^-}}
  f^\dag_{n\s'} c^\pd_{\beta\s'} \\ & \qquad
+ t^*_{\alpha m} t_{\beta n}
  f^\dag_{m\s} c^\pd_{\alpha \s} \frac{1}{E_{a\s} - E_{2e^-}}
  c^\dag_{\beta\s'} f^\pd_{n\s'}
\Big\}.
\end{align*}
which can be rewritten as
\begin{align}
  H_\mathrm{int} &= \sum_{\substack{\alpha, \beta; n, m \\ \s, \s'}} \left\{
         \frac{ t_{\alpha m} t_{\beta n}^* }{\e_n + 3 U}
       - \frac{ t_{\beta n}^* t_{\alpha m} }{\e_m + 2 U}
\right\}
         c_{\alpha\s}^\dag c_{\beta\s'}^\pd f_{n\s'}^\dag f_{m\s} \nonumber
\\ &- \sum_{\substack{\alpha, \beta; n, m \\ \s, \s'}}
         \delta_{n, m} \delta_{\s\s'}
         \frac{ t_{\alpha m} t_{\beta n}^* }{\e_n + 3 U}
         c_{\alpha\s}^\dag c_{\beta\s'}^\pd
\end{align}

The Kondo exchange interaction for the levels $n,m = \{1, 2, 3\}$
and thus for the orbital states $a,b = \{ s, h, p \}$
can straightforwardly be read off
this expression\begin{align*}
 J_{\alpha\beta}^{11} &= J_{\alpha\beta}^{hh}
 = 2 \Big( \frac{ t_{\alpha 1} t_{\beta 1}^* }{\e_1 + 3 U}
         - \frac{ t_{\beta 1}^* t_{\alpha 1} }{\e_1 + 2 U} \Big)
  \\
 J_{\alpha\beta}^{22} &= J_{\alpha\beta}^{ss}
 = 2 \Big( \frac{ t_{\alpha 2} t_{\beta 2}^* }{\e_2 + 3 U}
         - \frac{ t_{\beta 2}^* t_{\alpha 2} }{\e_2 + 2 U} \Big)
  \\
 J_{\alpha\beta}^{33} &= J_{\alpha\beta}^{pp}
 = 2 \Big( \frac{ t_{\alpha 3} t_{\beta 3}^* }{\e_3 + 3 U}
         - \frac{ t_{\beta 3}^* t_{\alpha 3} }{\e_3 + 2 U} \Big)
  \\
 J_{\alpha\beta}^{12} &= J_{\alpha\beta}^{sh}
 = 2 \Big( \frac{ t_{\alpha 2} t_{\beta 1}^* }{\e_1 + 3 U}
         - \frac{ t_{\beta 1}^* t_{\alpha 2} }{\e_2 + 2 U}  \Big)
  \\
 J_{\alpha\beta}^{21} &= J_{\alpha\beta}^{hs}
 = 2 \Big( \frac{ t_{\alpha 1} t_{\beta 2}^* }{\e_2 + 3 U}
         - \frac{ t_{\beta 2}^* t_{\alpha 1} }{\e_1 + 2 U}  \Big)
 \\
 J_{\alpha\beta}^{32} &= J_{\alpha\beta}^{ps}
 = 2 \Big( \frac{ t_{\alpha 2} t_{\beta 3}^* }{\e_3 + 3 U}
         - \frac{ t_{\beta 3}^* t_{\alpha 2} }{\e_2 + 2 U}  \Big)
  \\
 J_{\alpha\beta}^{23} &= J_{\alpha\beta}^{sp}
 = 2 \Big( \frac{ t_{\alpha 3} t_{\beta 2}^* }{\e_2 + 3 U}
         - \frac{ t_{\beta 2}^* t_{\alpha 3} }{\e_3 + 2 U}  \Big)
\end{align*}
which leads to the interaction Hamiltonian as defined in the main text.
Note that the potential scattering which changes the orbital index
is of the same strength as the spin Kondo scattering with an orbital
change, $J^{h/p, s}_{\alpha\beta}$.
In lowest order, i.e.~involving two hopping processes,
there is no transition between the hole and the particle state
and we assume for the following
\begin{align*}
  J_{\alpha\beta}^{hp} =   J_{\alpha\beta}^{ph}  = 0.
\end{align*}

Furthermore we disregard a constant shift due to the
following potential scattering contribution
\begin{align*}
  H_{pot.scat} =& - \eh \sum\limits_{\alpha\beta} {\cal C}_{\alpha\beta}
  \Big( \sum_\s c_{\alpha \s}^\dag c_{\beta \s}  \Big)
\end{align*}
since it is negligible small at the particle-hole symmetric point
$\e_2 = - 5/2\ U$,
\begin{align*}
{\cal C}_{\alpha\beta}
=& \Big( \frac{t_{\alpha 1} t_{\beta 1}^*}{\e_1 + 3 U}
               +  \frac{t_{\beta 1} t_{\alpha 1}^*}{\e_1 + 2 U}  \Big)
\\ &
        + \Big( \frac{t_{\alpha 2} t_{\beta 2}^*}{\e_2 + 3 U}
                + \frac{t_{\beta 2} t_{\alpha 2}^*}{\e_2 + 2 U} \Big)
        + \Big( \frac{t_{\alpha 3} t_{\beta 3}^*}{\e_3 + 3 U}
                    + \frac{t_{\alpha 3} t_{\beta 3}^*}{\e_3 + 2 U} \Big)
\\
=& t_{\alpha 1} t_{\beta 1}^*\frac{2 \delta}{(U/2)^2 - \delta^2}
 - t_{\alpha 3} t_{\beta 3}^*\frac{2 \delta'}{(U/2)^2 - (\delta')^2}.
\end{align*}
This constant offset in the current of order $\delta/U^2$
is neglected in our calculation.

\section{Calculation of the Self~Energies}\label{app:selfen}

The Green's functions for the orbital states $a, b = \{ s, h, p \}$
are calculated by an expansion in the
interaction Hamiltonian
\begin{align}
&\mathcal{G}_{ab,\s\s'}(\tau,\tau^{\prime})
=-\mathrm{i}\left\langle T_{C_K}
d_{a\s}^\pd(\tau)d_{b\s'}^{\dagger}(\tau^{\prime})\right\rangle_0 \nonumber\\
&+ (-\mathrm{i})^{2}\int_{C_K}\mathrm{d}\tau_{1}\left\langle T_{C_K}
d_{a\s}^\pd(\tau)H_\mathrm{int}(\tau_{1})
d_{b\s'}^{\dagger}(\tau^{\prime})\right\rangle_0 \nonumber\\
&+
\frac{(-\mathrm{i})^{3}}{2}\int_{C_K}\mathrm{d}\tau_{1}\mathrm{d}\tau_{2}
\left\langle
T_{C_K}d_{a\s}^\pd(\tau)H_\mathrm{int}(\tau_{1})H_\mathrm{int}(\tau_{2})
d_{b\s'}^{\dagger}(\tau^{\prime})\right\rangle_0 \nonumber\\
&+ \mathcal{O}(H_\mathrm{int}^{3})\label{series}
\end{align}
The linear order does not contribute since it contains the
expectation value of $\langle \mathbf{s}_{\alpha\beta}\rangle$ which
is zero if the spin in the leads are not polarized. In the case of
ferromagnetic leads this order has to be taken into account, but in
our setup the leading order is second order in the coupling.

We get contributions from the spin part of the interaction
Hamiltonian and of the potential scattering part. A mixing between
both parts does not appear in second order due to the convolution of
the leads contribution.

For example the spin part is given by (Einstein's sum rule)
\begin{align}
\Gc&_{ab,\s\s'}^{(2), \mathrm{Spin}}(\tau, \tau') 
= \frac{\ii}{2} \int \mathrm{d}\tau_1 \mathrm{d}\tau_2
\ev \boldsymbol{\tau}^{i_1}_{\s_1'\s_1} \boldsymbol{\tau}^{i_2}_{\s'_2\s_2}
\nonumber \\ &\qquad \times
\left\langle T_\mathrm{C_K}
c^\dag_{\alpha \s'_1}(\tau_1) c^\pd_{\beta \s_1}(\tau_1)
c^\dag_{\alpha' \s'_2}(\tau_2) c^\pd_{\beta' \s_2}(\tau_2)
\right\rangle_0\nonumber \\
&\qquad\times \left\langle T_\mathrm{C_K} d^\pd_{a\s}(\tau)
J_{\alpha\beta}^{a_1 b_1} \mathbf{S}^{i_1}_{a_1 b_1}(\tau_1)
J_{\alpha'\beta'}^{a_2 b_2} \mathbf{S}^{i_2}_{a_2 b_2}(\tau_2)
d_{b\s'}^\dagger(\tau') \right\rangle_0 \nonumber
\end{align}
The conduction electron spins contract to
\begin{align}
\boldsymbol{\tau}_{\s' \s} \boldsymbol{\tau}_{\s \s'}
\delta_{\alpha'\beta} \delta_{\alpha\beta'}
G_{\beta\sigma'}^{(0)}(\tau_1,\tau_2)
G_{\alpha\s}^{(0)}(\tau_2,\tau_1).\nonumber
\end{align}
Since the leads are not magnetic, the conduction electron Green's
functions does not depend on spin and the sum over the spins acts
only on the $\tau$-matrices and we can thus introduce
the conduction electron spin susceptibility
$\chi_{\alpha\beta}(\tau_2, \tau_1) =
G_{\alpha}^{(0)}(\tau_2,\tau_1) G_{\beta}^{(0)}(\tau_1,\tau_2)$.

As discussed before there can in general be solutions with $a \not = b$
which we neglect in this setup~\cite{Koerting:08}.
The result for the diagonal Greens's function yields finally the second order
self energy by comparison with the Dyson series:
\begin{align*}
  \Sigma_{a\s}(\tau_1, \tau_2)
&= \frac{1}{16} \Big( J_{\alpha\beta}^{ab} J_{\beta\alpha}^{ba}
                    + J_{\alpha\beta}^{ba} J_{\beta\alpha}^{ab}
                \Big) \chi_{\alpha\beta}(\tau_2,  \tau_1)
\\ & \quad
   \big[ 2 \boldsymbol{\tau}_{\s\s'} \boldsymbol{\tau}_{\s'\s}
      + \delta_{\s,\s'} {\cal P}_{a,b} \big]
         G_{b\s'}(\tau_1, \tau_2) 
\\ &
  + \Sigma_{a\s}^\mathrm{el}(\tau_1, \tau_2)
\end{align*}
where the inelastic and elastic 
potential scattering contributions are defined as
\begin{align*}
{\cal P}_{ab} &=
    \big( \delta_{a, s}\delta_{b, p} + \delta_{b,s}\delta_{a, p}
        + \delta_{a, s} \delta_{b, h} + \delta_{b, s} \delta_{a, h} \big)
\\
\Sigma_{a\s}^\mathrm{el}(\tau_1, \tau_2)
 &= \frac{1}{8}  \sum_{\tilde a}
                J_{\alpha\beta}^{\tilde a \tilde a} 
                J_{\beta\alpha}^{\tilde a \tilde a}
                \chi_{\alpha\beta}(\tau_2,  \tau_1)
\\ & 
  \big( \delta_{\tilde a, p} \delta_{a, h} + \delta_{\tilde a, s} \delta_{a, p}
      + \delta_{\tilde a, p} \delta_{a, s}
  \big)  
                G_{a \s}(\tau_1, \tau_2)
\end{align*}
Note that in the rate equation $\Sigma_{a\s}^{el}$ does not
contribute since it does not contain transitions between states.

The self-energy components needed in Eq.~\eqref{sce} are now readily
obtained from analytical continuation using the Langreth rules.

\end{appendix}


\end{document}